\documentstyle[prl,rotate,amssymb,psfig,eqsecnum,aps]{revtex}
\input{epsf}

\begin{document}
\wideabs{
\title{Observation of an Anomalous Number of Dimuon Events in a High Energy Neutrino Beam }
\author{T.~Adams$^{4}$, A.~Alton$^{4}$, S.~Avvakumov$^{8}$, L.~de~Barbaro$^{5}$,
P.~de~Barbaro$^{8}$, R.~H.~Bernstein$^{3}$, A.~Bodek$^{8}$, T.~Bolton$^{4}$,
J.~Brau$^{6}$, D.~Buchholz$^{5}$, H.~Budd$^{8}$, L.~Bugel$^{3}$, 
J.~Conrad$^{2}$, R.~B.~Drucker$^{6}$, B.~T.~Fleming$^{2}$, R.~Frey$^{6}$, 
J.A.~Formaggio$^{2}$, J.~Goldman$^{4}$, M.~Goncharov$^{4}$, D.~A.~Harris$^{8}$,
R.~A.~Johnson$^{1}$, J.~H.~Kim$^{2}$, S.~Koutsoliotas$^{2}$, 
M.~J.~Lamm$^{3}$, W.~Marsh$^{3}$, D.~Mason$^{6}$, J.~McDonald$^{7}$, 
C.~McNulty$^{2}$,
K.~S.~McFarland$^{3}$, D.~Naples$^{7}$, P.~Nienaber$^{3}$, A.~Romosan$^{2}$,
W.~K.~Sakumoto$^{8}$, H.~Schellman$^{5}$, M.~H.~Shaevitz$^{2}$,
P.~Spentzouris$^{2}$, E.~G.~Stern$^{2}$, N.~Suwonjandee$^{1}$, M.~Tzanov$^{7}
$, M.~Vakili$^{1}$, A.~Vaitaitis$^{2}$, U.~K.~Yang$^{8}$, J.~Yu$^{3}$,
G.~P.~Zeller$^{5}$, and E.~D.~Zimmerman$^{2}$}
\date{\today}
\address{
$^1$University of Cincinnati, Cincinnati, OH 45221 \\
$^2$Columbia University, New York, NY 10027 \\
$^3$Fermi National Accelerator Laboratory, Batavia, IL 60510 \\
$^4$Kansas State University, Manhattan, KS 66506 \\
$^5$Northwestern University, Evanston, IL 60208 \\
$^6$University of Oregon, Eugene, OR 97403 \\
$^7$University of Pittsburgh, Pittsburgh, PA 15260 \\
$^8$University of Rochester, Rochester, NY 14627 \\ 
}
\maketitle

\begin{abstract}
A search for long-lived neutral particles ($N^{0}$'s) with masses above 2.2 
GeV$/c^{2}$ that decay into at least one muon has been performed using an
instrumented decay channel at the NuTeV experiment at Fermilab. Data were
examined for particles decaying into the final states $\mu \mu $, $\mu e$,
and $\mu \pi $. Three $\mu \mu $ events were observed over an expected
Standard Model background of $0.069 \pm 0.010$ events; no events 
were observed in the other modes. 

\vspace{0.1in} \noindent . PACS numbers: 14.80.-j, 13.85.Rm, 12.60.Jv, 13.15+g
\end{abstract}
\twocolumn
}

%\draft

%\twocolumn[\maketitle]

%\bigskip 

%{\normalsize \newpage }%\narrowtext

%\pssilent

A number of extensions to the Standard Model (SM) of elementary particle
physics predict new particles with reduced couplings to normal quarks and
leptons. Some of these particles, which we shall refer to as $N^{0}$'s, may
have zero electric charge, long lifetimes, and small interaction rates with
ordinary matter. Examples include neutral heavy leptons (NHLs) or heavy
sterile neutrinos~\cite{grl,shrock,tim} and neutral supersymmetric 
particles~\cite{borissov_susy}. The $N^{0}$ may be either pair-produced 
in hadronic
interactions or occur as a daughter particle in weak decays of mesons
through mixing with SM neutrinos. Decays of the $N^{0}$ can proceed through 
$W^{\pm }$ or $Z^{0}$ mediated decays via mixing, or through $R$-parity
violating supersymmetric processes. \ We report here the results of a search 
for $N^{0}$ particles in the mass region above $2.2$ GeV/$c^{2}$ that decay into
final states with at least one muon and one other charged particle, using the
NuTeV detector at Fermilab (E815). This analysis extends our results for a  
previous search for $N^{0}$'s in the mass region 
$0.3$ to $3.0$ GeV/$c^{2}$ with at least one final
state muon~\cite{arturv_prl}, and for a search for $N^{0}$'s with mass $\leq$ 
$0.3$ GeV/$c^{2}$ that decay to $e^{+}e^{-}\nu$~\cite{karmino_prl}.

NuTeV received $2.54\times 10^{18}$ 800~GeV/$c$ protons during the 1997
fixed-target run at Fermilab with the detector configured for this search.
The proton beam was incident on a one-interaction-length beryllium oxide
target at a targeting angle of 7.8 mr with respect to the detector. A
sign-selected quadrupole train (SSQT)~\cite{SSQT} focused either 
positive (45$\%$ of the running) or negative (55$\%$ of the running) 
secondary $\pi $ and $K$ mesons into a 440 m evacuated pipe pointed
toward the NuTeV experiment. Surviving neutrinos, and possibly  $N^{0}$'s,
traversed $\sim $850 meters of earth-steel shielding before reaching the
NuTeV decay channel.

\begin{figure}
 \unitlength1cm
  \centerline{\psfig{figure=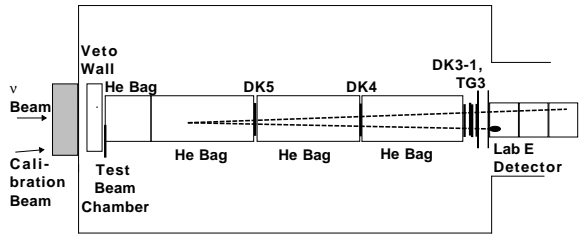,width=0.45\textwidth,clip=T}}
\caption{Schematic of the NuTeV decay channel with example $N^0 \rightarrow
\mu \pi$ decay.}
\label{fig:dkchannel}
\end{figure}

The decay channel region (Fig.~\ref{fig:dkchannel}), located 1.4 km
downstream of the production target, was designed to search for exotic
neutral particle decays in a $\left( 1.27\times 1.27\times 34\right) $~m$^{3}
$ fiducial volume. The volume consisted of large helium-filled plastic bags
interspersed with drift chambers (DC). An array of plastic scintillation
counters at the upstream end of the decay channel provided a means to veto
charged particles produced in the upstream shielding. The Lab E neutrino
detector~\cite{detector,nim} downstream of the decay channel provided final
state particle energy measurement and identification. This detector
consisted of a 690 ton iron-scintillator sampling calorimeter, instrumented
with DC for charged particle tracking, followed by a toroid muon
spectrometer. Sets of hits in the calorimeter DC were linked to tracks found
with the decay channel DC system to determine particle identification. Muons
were identified by penetration, electrons by their characteristic short
clusters of hits, and charged hadrons by their elongated hit patterns. All
charged hadrons were assumed to be pions. The particle identification
algorithm was tuned with electrons, pions, and muons provided by the NuTeV
calibration beam~\cite{nim}. The probability to mis-reconstruct a 
dimuon event as a $\mu e$ or $\mu\pi$ event was 0.2\%; 
$\mu e$  and $\mu\pi$ events were mis-reconstructed
as $\mu\mu$ events with a probability of 0.4\% and
1.3\% respectively.  Muon energy was determined by either
spectrometer measurement ($\sigma _{p}/p=11\%$), range in the calorimeter 
($\sigma _{p}=310$ MeV), or multiple scattering ($\sigma _{p}/p=42\%$ at  
$p=50$ GeV/$c$). Only muons penetrating to the toroid spectrometer could be
charge-identified.  The hadronic energy resolution of the calorimeter was
$\sigma/E=(0.024\pm 0.001)\oplus(0.874\pm0.003)/\sqrt{E{\rm (GeV)}}$;
the electromagnetic energy resolution,
${\sigma }/{E} =(0.04\pm 0.001)\oplus{(0.52\pm 0.01)}/{\sqrt{E{\rm (GeV)}}}$~\cite{nim}.

Event selection criteria were developed to minimize known backgrounds while
maintaining efficiency for a possible $N^{0}$ signal. \ A series of cuts
isolated events with exactly two well-reconstructed tracks forming a vertex
within the decay channel fiducial volume and having no charged particle
identified in the upstream veto system. Both tracks were required to be
well-reconstructed and have an associated calorimeter cluster, with at least
one of the tracks identified as a muon. The track and vertex quality 
criteria were numerically the same as in Ref.~\cite{arturv_prl}.  The
probability for a signal event to fail this is
$<$2\%. The vertex position was required to be within the detector
fiducial volume; in addition, the longitudinal distance from the
vertex position to any drift chamber was required to be greater than 
the larger of 101.6~cm and $3\sigma _{z}$, with $\sigma _{z}$ the
longitudinal vertex position error. A third track which formed a downstream
vertex with one of the two initial tracks was permitted, to allow for 
$\delta$-ray emission.  Cosmic ray tracks (which are suppressed by
the fast gate timing of the neutrino beam) were removed by requiring 
the slope of each track relative to the
beam direction be less than 100~mr. Muons, hadrons, and electrons were
required to have an energy greater than 2.2~GeV, 10~GeV, and 10~GeV,
respectively, with an additional total energy cut of 12 GeV applied to $\mu
\mu $ events. In order to isolate high mass events, a transverse mass cut 
$m_{T}>2.2$~GeV/$c^{2}$ was applied, with $m_{T}\equiv |P_{T}|+\sqrt{
P_{T}^{2}+m_{V}^{2}}$, $P_{T}$ the component of the total reconstructed
momentum perpendicular to the beam direction, and $m_{V}$ the invariant mass
of the visible particles.

Further ``clean event'' cuts were applied to reduce the level of the
dominant deep-inelastic neutrino scattering (DIS) backgrounds. DIS events
typically had large track multiplicities, many drift chamber hits, and
extra unassociated clusters in the calorimeter. Clean cuts required: (1)
three or fewer tracks in any one DC view, (2) three or fewer DC hits in any
view of the first chamber downstream of the vertex, (3) at least one DC view
with fewer than eight hits total in the first two chambers downstream of
the vertex, (4) no energy clusters in the calorimeter not associated with
tracks, and (5) no tracks identified as electrons with missing hits in
either view of the first two chambers downstream of the vertex. The final
cut removed events where a photon from the primary vertex converted to $e^+e^-$
and was reconstructed as an electron.

Detailed MC simulations of physics processes including detector effects were
used to estimate possible  backgrounds to the $N^{0}$ signal.  Two
major classes of physics
processes considered included DIS~\cite{seligman}, resonance production~\cite
{lownu}, and diffractive scattering~\cite{diffnutev} by neutrinos; and decays
and interactions of hadrons and photons produced in neutrino interactions.
Particular attention was given to known sources of dimuon or dimuon-like
production: DIS, resonance, and diffractive production of charm;
neutrino trident production; $\mu^{+}\mu ^{-}$
vector meson decays; electromagnetic muon pair production; low multiplicity 
$\nu _{\mu }$ DIS\ accompanied by a secondary pion or kaon decay;  and decays
of $K_{L}^{0}$ mesons produced by neutrino interactions in the decay channel
or surrounding material.   Approximately $10^3$ times as many neutrino
interactions as in the data were simulated in the decay channel volume;
in addition, a large sample of DIS\ events was generated in the material
surrounding the decay channel.  Event generators
fed a GEANT-based~\cite{geant} detector simulation that produced hit-level
simulations of raw data including DC inefficiencies and noise. MC events
were processed using the same analysis routines used for the data. A number
of other possible sources were studied as well and found to be 
negligible.

\begin{figure}
\centerline{\psfig{figure=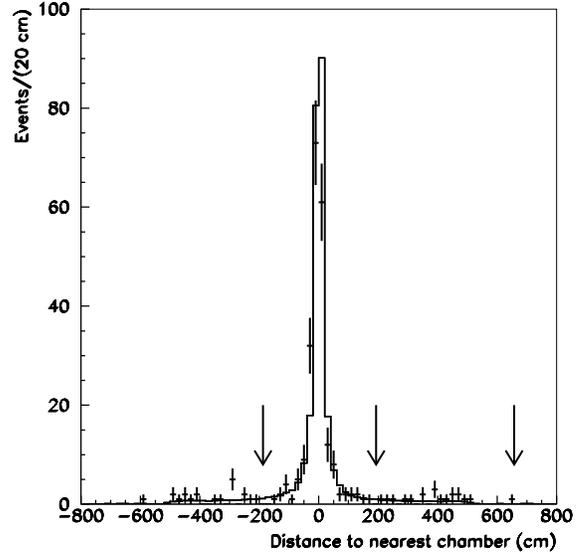,width=0.42\textwidth,clip=T}}
\caption{ Distance of longitudinal vertex position from the closest chamber for
events in the normalization sample without the $z$ vertex
requirement. (Crosses: data; histogram: MC.; 
arrows: values for the three observed $\mu\mu$ events)}
\label{nearch}
\end{figure}

Background calculations were normalized to data using charged-current DIS
interactions in the decay channel DC. Events in this sample were required to
pass the following five normalization cuts: a vertex within the transverse
fiducial volume ; a $z$ vertex within 76.2 cm of a DC; no upstream veto; 
$\geq $1~GeV energy deposit in the front of the calorimeter; and one
toroid-analyzed muon matched to a decay channel track. The MC was normalized
to match the total number of data events with two or more tracks; the error
on this normalization is $9\%$. 
Figure~\ref{nearch} shows a comparison of 
the longitudinal vertex resolution between data and MC with the $z$
vertex requirement removed.

MC events were also compared to another data control sample as a check on
the quality of the simulation. For this sample, the vertex was required to
be within the decay channel transverse fiducial volume but the $z$ position
was allowed to be either in the chambers or the helium. Tight track angle
cuts were imposed, and there was a strict requirement on veto system
activity. The majority of these events were from interactions in the chamber
material or from interactions in the laboratory floor. Of 495 events in
the data, 116 had vertices reconstructed in the helium at least 
$101.6$~cm from the nearest DC. This can be compared to the MC, which predicted 
$(514\pm 82)$ total events and $(96\pm 15)$ events reconstructed in the
helium. Because loose vertex quality requirements allowed 
mis-reconstructed interactions in the chambers and floor to enter this 
sample, only 35\% of the vertices reconstructed in the helium were actually
due to $\nu$-He interactions. 

After all cuts, the expected background to the $N^0$ search
was 0.069 $\pm $ 0.010 events in $\mu \mu $ mode, 0.13 $\pm $
0.02 events in $\mu e$ mode and 0.14 $\pm $ 0.02 events in $\mu \pi 
$ mode. Table~\ref{backg} summarizes the background components for
$\mu \mu$ mode; DIS clearly dominates. 

Before looking at the data in the signal region, we performed a series of
analyses on other fiducial and kinematic ranges. These included using: 
(1)~identical analysis cuts applied to events within 15.2 cm of a DC (the
``chamber region''); (2)~the chamber region with loosened cuts to increase 
$\mu \pi $ acceptance; (3)~the ``intermediate region'' between 15.2 and 
101.6~cm from the chambers, with otherwise standard analysis cuts; and 
(4)~events with well-reconstructed two-track vertices where the tracks were 
both identified as pions. Measurements agreed with MC predictions within 
$1.5\sigma$ in all cases. For example in the ``chamber region'' sample
(1), two two-track events were observed in the data with
3.5 predicted by the MC.

Upon examining the signal region, three $\mu \mu
(\nu )$ events were observed, which is considerably above the predicted
background. No $\mu e$ or $\mu \pi $ events were observed, which is
consistent with background estimates. Table~\ref{evt_kine} summarizes event
reconstruction characteristics of the $\mu \mu $ events.  The vertex and
track probabilities have been measured with both the signal and background
Monte Carlos and represent the probability that the vertex (or track)
would reconstruct with a similar (or worse) quality.

Observation of three $\mu \mu $ data events prompted further tests comparing
data to MC predictions with relaxed cuts. These included gradually removing
the cuts and alternately releasing and then returning individual cuts and
cut-pairs. These studies revealed no significant discrepancy between data
and the background calculations (except for the original 3 $\mu \mu $
events).

The three $\mu \mu $ events have some features consistent with a $N^{0}$
decay hypothesis. The events pass the analysis cuts, where the background is
estimated to be 0.069 events. All three occur well within the fiducial volume
away from the chambers and are evenly distributed throughout the decay
channel. The transverse mass, invariant mass, and missing $P_{T}$ are all
consistent with the decay of an $N^{0}$ with a mass of about
5 GeV/$c^2$.  Since
only 1(0) $\mu e$ and 2(0) $\mu\pi$ events (consistent with MC
expectations) were observed in the chamber (helium) data, it is unlikely
that $\mu\mu$ events are related to low multiplicity neutrino events
followed by $\pi$ and $K$ decay.

Unlike the background, in both an NHL or neutralino model one would expect
the $\mu \pi $ rate to be highly suppressed relative to leptonic decays.
However, for a 5 GeV/$c^{2}$ NHL model, one would expect 1.4 times more $\mu
e\nu $ events~\cite{tim}. A neutralino model, on the other hand, can
accommodate the observation of either only $\mu \mu $ or a combination of 
$\mu\mu$ and $\mu e$ candidates by selecting appropriate couplings.

However, several aspects of the candidate events are similar to those from
neutrino interaction backgrounds, and might be indicative of unaccounted-for
sources or a statistical fluctuation. Globally, the events share one feature
that is improbable for an $N^{0}$ decay hypothesis. All three events have a
muon energy asymmetry $A>0.83$, where $A\equiv \left( |E_{1}-E_{2}|\right)
/\left( E_{1}+E_{2}\right) $. For DIS background, the probability for three 
$\mu \mu $ events which pass the signal cuts to have the observed energy
asymmetry is $38\%$. The probability that this occurs in a weak decay
hypothesis~\cite{formaggio} is less than $0.5\%$ (including acceptance). All
three events occurred during the higher rate $\nu $-mode rather than 
$\overline{\nu }$-mode running periods. In the two events where the charge of
the higher-momentum muon can be measured, it has the same sign as expected
for the leading muon in a charged-current neutrino interaction. 
Event kinematics 
($M_{T}$, $M_{\mu \mu }$, $P_{T}$) are also consistent with DIS\
characteristics.  However, the observed number of events is inconsistent
with expected neutrino interaction background.

\begin{figure}
  \unitlength1cm
 \begin{picture}(9.0,6.0)(0,0)
  \put(0.3,0.5){\psfig{figure=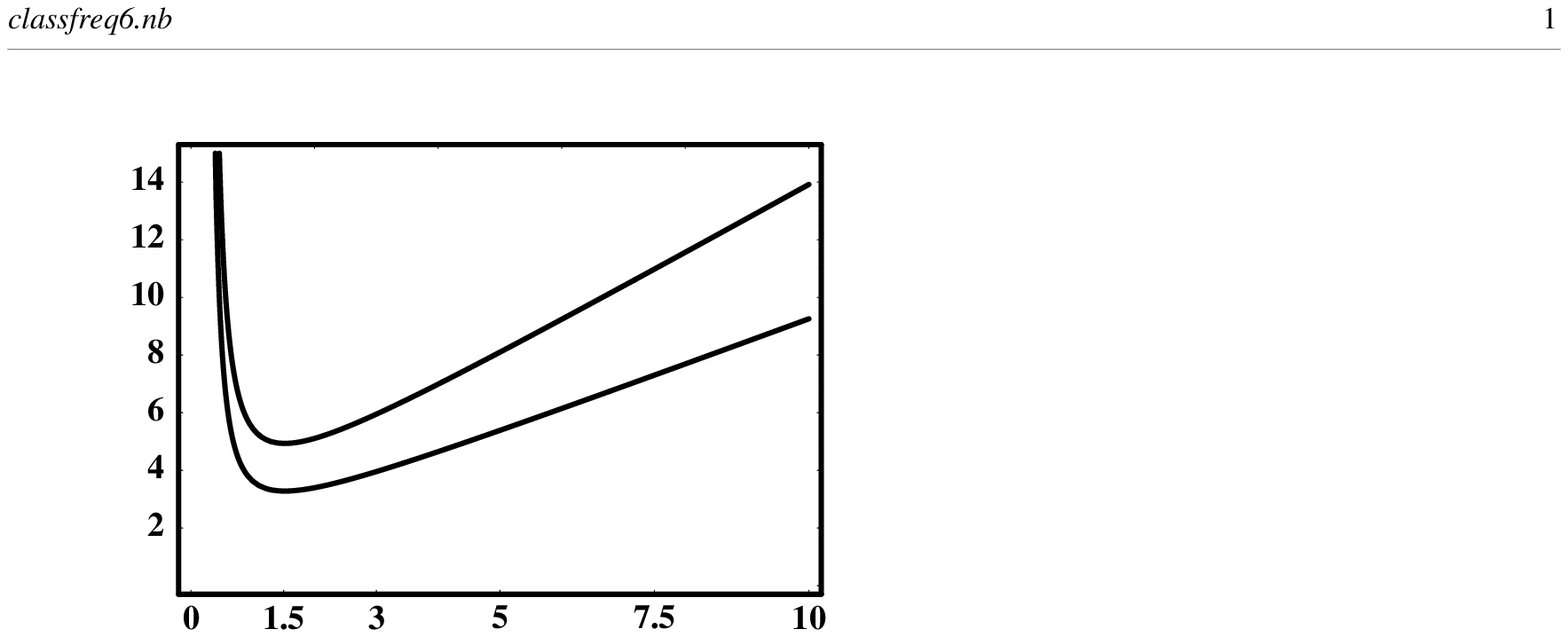,width=0.45\textwidth,clip=T}}
  \put(5.2,0.3){Decay length (km)}
  \put(0.22,2.9){\rotate[l]{d$\sigma$/d$\Omega$ (10$^{-11}$ mb)}}
  \put(4.3,4.0){99\%}
  \put(5.5,2.7){90\%}
 \end{picture}
 \caption{NuTeV limit on neutralino production. The
plot gives the limit on the differential p-p cross section to produce an
$N^0$ versus the decay length of the $N^0$ in the laboratory
coordinate system.  This limit is generic for
   an $N^0$ produced at the target.}
 \label{neutlim}
\end{figure}

In summary, NuTeV has observed three $\mu \mu $ events, zero $\mu \pi $, and
zero $\mu e$ events with transverse mass above 2.2 GeV/$c^{2}$. The expected
backgrounds were $0.069\pm 0.010$, $0.14\pm 0.02$, and 
$0.13\pm 0.02$ events, respectively. The rate
corresponding to the observed three events is not consistent with Standard
Model processes we have identified and the source of the events is not
clear. 
NuTeV is insensitive to NHL production in this mass region, but can
set an interesting limit on neutralino production 
(determined by calculating
one-sided limits using a frequentist approach without background
subtraction~\cite{feldman}). NuTeV is the first experiment to set 
limits on the production
of long-lived neutralinos in this mass range which decay by $R$-parity
violation. This limit (Fig.~\ref{neutlim}), although motivated by a
neutralino hypothesis, is a generic limit applicable for any model of
neutral particle production at the target~\cite{borissov_susy}.

This research was supported by the U.S. Department of Energy and the
National Science Foundation. We thank the staff of FNAL for their
contributions to the construction and support of this experiment during the
1996-97 fixed target run.

\bigskip

\bigskip

\newpage

\begin{table}
\caption{Estimated rates of background to the $N^0\rightarrow \mu \mu (\nu )$
search \label{backg}}
\begin{tabular}{lc}
Source & $\mu \mu (\nu )$ events \\ \hline
DIS events        & (6.8 $\pm $ 1.0) $\times $ 10$^{-2}$ \\ 
Diffractive charm & (1.3 $\pm $ 0.1) $\times $ 10$^{-3}$\\ 
Diffractive $\pi$ & (1.9 $\pm $ 0.1) $\times $ 10$^{-4}$ \\ 
Diffractive $K$   & (4.0 $\pm $ 0.3) $\times $ 10$^{-7}$ \\ 
K$_{L}^{0}$ decays from shielding & (3.9 $\pm $ 3.9) $\times $ 10$^{-4}$ \\ 
Other sources & $ \ll$ 2.5 $\times$ 10$^{-4}$ \\
\hline
Total $\mu \mu (\nu )$ Background & (6.9 $\pm $ 1.0) $\times $ 10$^{-2}$ \\ 
\end{tabular}
\end{table}

\begin{table}
\caption{Kinematic and reconstruction quantities for the three 
candidate $N^0\rightarrow \mu\mu(\nu)$ events. The variables refer
to the muons energies (E$_{\mu 1}$, E$_{\mu 2}$), missing transverse
momentum ($P_{T{\rm miss}}$), two muon invariant mass ($m_{\rm inv}$),
transverse mass ($m_T$, see text), transverse vertex position ($v_x, v_y$),
longitudinal distance to nearest drift chamber ($|\Delta z|$), vertex
probability (${\mathcal P}_{vert}$), first muon track probability 
(${\mathcal P}_{\mu 1}$), and second muon track
probability (${\mathcal P}_{\mu 2}$). The probabilities are measured
the with signal(background) Monte Carlos.  The sign on the 
muon energy refers to the charge of the muon (if measured). \label{evt_kine}} 
\begin{tabular}{cccccccccc}
 Event                       &      1     &       2     &      3      \\ \hline
 E$_{\mu 1}$ (GeV)           &   --77.4   &    --92.0   &  $\pm $48.0 \\
 E$_{\mu 2}$ (GeV)           & $\pm 2.56$ &  $\pm 5.85$ &  $\pm 4.34$ \\
 $P_{T{\rm miss}}$ (GeV/$c$) &    2.42    &     1.41    &     2.07    \\
 $m_{\rm inv}$ (GeV/$c^2$)   &    1.10    &     0.88    &     3.57    \\
 $m_{T}$ (GeV/$c^2$)         &    5.08    &     3.08    &     4.66    \\
 $v_x$ $(cm)$                &   --46.5   &     48.0    &    --57.5   \\
 $v_y$ $(cm)$                &    3.4     &    --38.3   &     15.0    \\
 $|\Delta z|$ $(cm)$         &    193     &      657    &    --188    \\ 
 (${\mathcal P}_{vert}$)     & 0.81(0.94) & 0.043(0.48) & 0.011(0.30) \\
 (${\mathcal P}_{\mu 1}$)    & 0.95(0.96) & 0.034(0.18) & 0.51(0.73)  \\
 (${\mathcal P}_{\mu 2}$)    & 0.49(0.70) & 0.034(0.18) & 0.15(0.39)  \\
\end{tabular}
\end{table}

\end{document}